# Rocket Exhaust Blowing Soil in Near Vacuum Conditions is Faster than Predicted by Continuum Scaling Laws

Philip T. Metzger[1]

[1] Florida Space Institute, University of Central Florida, 12354 Research Parkway, Partnership 1 Building, Suite 214, Orlando, FL 32826-0650; PH (407) 823-5540; email: philip.metzger@ucf.edu

## ABSTRACT

Experiments were performed to study how rocket exhaust blows soil in lunar and Martian conditions. Jets of gas were blown downwardly at various granular materials while a camera recorded the formation of scour holes as the material was removed. The experiments were performed in a series of conditions ranging from ambient 101 kPa pressure down to 0.3 kPa. This includes the range that is relevant for lunar conditions, because it is not hard vacuum inside the rocket exhaust where soil is being eroded. Prior work in ambient conditions showed that erosion rate is proportional to the square of the densimetric Froude number. However, the prior work was not performed at low pressures. In the new work, preliminary results show that as the gas density is so low that gas flow around the sand grains is no longer in the continuum regime, then erosion rate is faster than predicted by the earlier results. The dependency of erosion rate on various parameters is determined, but it turns out to be more complex than expected so additional experimental work will be required to develop complete scaling relationships.

## EXPERIMENTAL SETUP

Tests were performed at the Planetary Aeolian Lab (PAL) at the NASA Ames Research Center. The PAL is inside a large-volume tower which is evacuated to partial vacuum by means of a steam ejector system, achieving pressures as low as 2 Torr (267 Pa). Ten boxes of sand were arranged in a circle in this vacuum chamber and a video camera in the center could pan to observe any one of the ten boxes at a time. Each box was open on the top and filled with a granular material. The front wall of the box facing the camera was clear plastic with a beveled top edge. The jet was blown onto the bevel so that the gas flow was split into two, half entering the granular material while remaining attached to the inside of the clear wall; the other half blowing outside the box. Thus, scour holes formed inside the box attached to the clear wall so that the experiment was essentially split in half by the wall, permitting us to see inside the scour hole crater as it formed and evolved. Bottles of compressed nitrogen or helium gas were connected through a pressure regulator into an accumulator bottle that vented without a regulator into the rest of experiment. The gas then flowed through mass flow meters, a manifold, and a bank of solenoid valves to route the gas into the correct one of ten experiment boxes. The gas traveled down a long, straight pipe onto the top edge of the beveled wall. Care was taken to ensure the pipes were centered on the bevel with a constant vertical gap above the top edge of the wall. The experimental setup is shown in Figure 1.

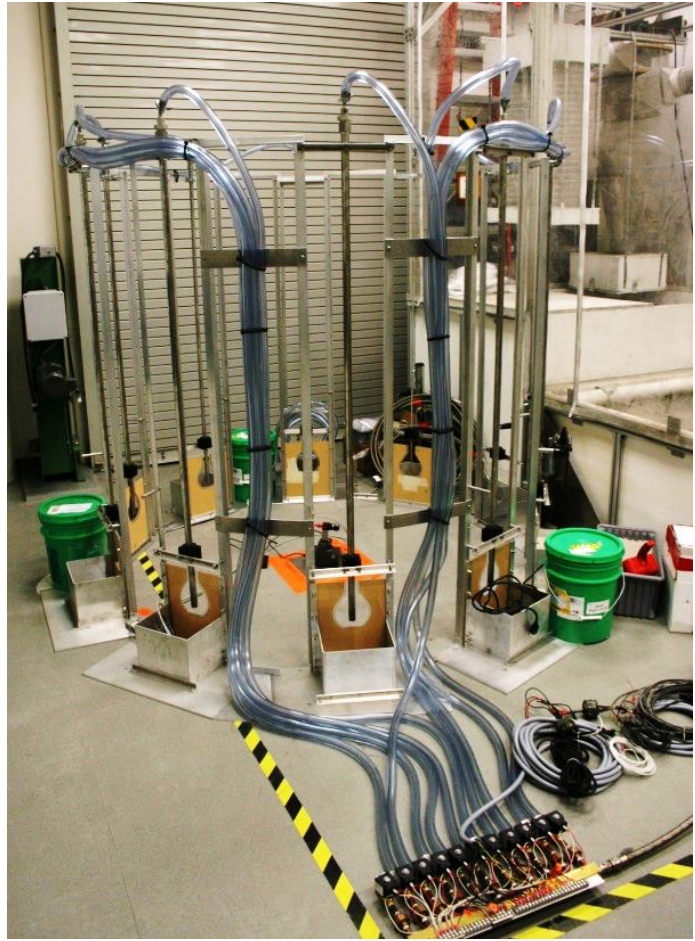

**Figure 1. Test Setup During Pre-Test Checkout at the Kennedy Space Center.**

**SCALING RELATIONSHIPS**

In prior work (Metzger et al, 2009a; Metzger et al, 2009b; Metzger et al, 2010) in the continuum regime the erosion rate was found to scale as follows:

$$\dot{V} = BA\frac{\rho v^2}{\rho_m g D}$$

where $\dot{V}$ is volumetric growth rate of the erosion crater, $B$ is a constant with units of velocity (m/s) whose physics are not yet understood (Metzger et al, 2010), $A$ is the area of the jet as it exits the pipe, $\rho_m$ is mineral density in the sand, $g$ is gravity, and $D$ is (average) sand grain diameter. Note the fraction in this equation is the square of the densimetric Froude number where the Archimedes buoyancy term is omitted since sand has negligible buoyancy in the gas. It can be understood as the kinetic energy density in the gas (energy to move the sand) divided by potential energy needed to lift a sand grain over its neighbor (energy to resist moving the sand). The numerator can also be seen as momentum flux, the rate that momentum enters the experiment, which says that erosion in the continuum regime is a momentum-driven process.

It was hoped that scaling in the transitional regime would take the form,

$$\dot{V} = BA \frac{\rho v^2}{\rho_m g D} \times f(Kn)$$

where $f$ is a function of only the Knudsen number $Kn$ that serves as a correction to the continuum scaling. The Knudsen number relative to a sand grain is defined as $Kn = \lambda/R$ where $R$ is radius of the sand grains (or average radius for the case of polydisperse granular materials), $\lambda = k_B T/(\sqrt{2}\pi d^2 P_0)$ is the mean free path length of the gas molecules in the jet, $k_B$ is Boltzmann's constant, $T$ is temperature of the jet, $d$ is collision diameter of the gas molecules and $P_0$ is the static pressure of the jet matching ambient pressure in the chamber. $Kn < 0.01$ is commonly defined as the continuum flow regime. $0.01 < Kn < 1.0$ is defined as the transitional flow regime. $1.0 < Kn$ is defined as the free molecular flow regime. It was hoped that the scaling relationship without $f$ would apply through the entire continuum regime and that $f$ could be derived for the transitional regime. The experiment cases were chosen to test this hypothesis and to derive $f$.

## VISUAL RESULTS

A total of 138 test cases were performed, including various granular materials (to vary grain size and mineral density), various pipe diameters, different gas velocities, and different vacuum levels (to vary Knudsen number). We were able to test into the transitional regime, but we could not achieve test cases into the free molecular flow regime because the chamber does not pump to lower pressures and because we could not use smaller sand grains without the material becoming cohesive, which would complicate the analysis.

An example of the initial results is shown in Figure 1. In this example, the vacuum chamber was pumped down to four different back ground pressures. The gas jets were exiting the pipes at subsonic velocities so their static pressure would be nearly equalized to the chamber pressure even before they exited the pipes, and this would determine the gas density, $\rho$. The jets were adjusted in velocity $v$ so that their dynamic pressure $P = \rho v^2 = 180.7$ Pa would remain constant in all four cases. Thus, velocity had to be set higher when chamber pressure was lower. The jets were applied to the sand for 120 seconds each time. The continuum scaling equation above says erosion rate is proportional to *P*. Thus, if the resulting crater volumes show a trend for these four cases where *P* is constant, then the rarefaction is having a nontrivial effect. These four cases were performed with these values of *Kn*: (A) 0.094, (B) 0.061, (C) 0.037, and (D) 0.009, with larger numbers indicating longer mean free path length in the gas, i.e., more rarefied conditions. Case (D) result is valid for Earth conditions where the Navier-Stokes equation is valid. Case (A) represents the conditions of a lunar or Martian rocket exhaust plume. It was found that the craters are monotonically larger with increasing *Kn*. This implies that the scaling relationships developed for continuum flow conditions under-predict the crater sizes for erosion in the transitional regime, which is relevant to lunar landings. This was an important finding.

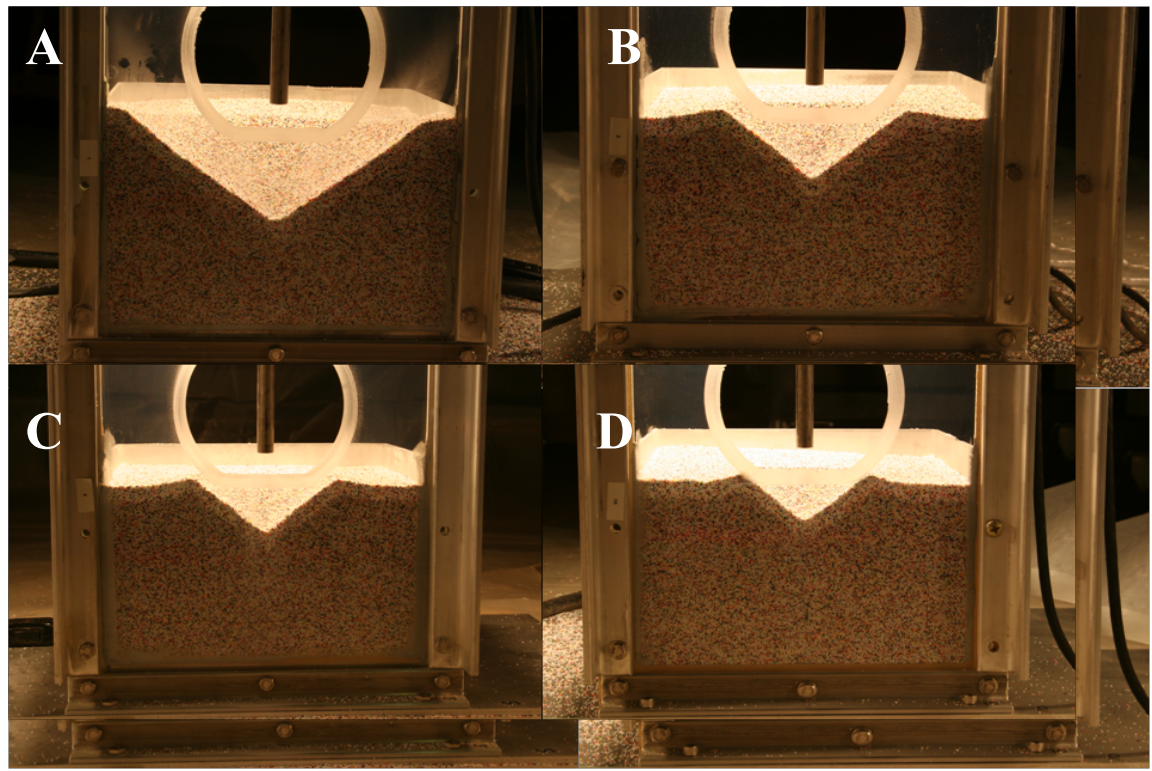

Figure 2: Test Cases with Varying Kundsen Number.

DETAILED RESULTS

The granular materials used in the tests are summarized in Table 1. The tests with play sand are summarized in Figure 3. These cases involved varying both $\rho$ and $v$ but all other parameters remained constant. The erosion rate was normalized by dividing with $A\rho v^2/(\rho_m g D)$. The erosion rate is approximately constant for *Kn*<0.01 (continuum regime) but increasing with *Kn* for 0.01<*Kn* (transitional regime). However, there is a large amount of scatter in the data suggesting that more variables than just *Kn* are affecting erosion rate in these cases.

Table 1. Granular Materials Used in Tests.

| Material | Mean Size (μm) | CU[1] | CC[1] | Permeability Kozeny & Carman (Darcies) | Material Density (g/cm$^3$) |
|---|---|---|---|---|---|
| "Play Sand" | 353.3 | 2.14 | 0.97 | 228 | 2.61 |
| Glass Beads | 214.8 | 1.37 | 1.15 | 31.9 | 2.44 |
| Plastic Beads | 798.2 | 1.25 | 1.06 | 1463 | 1.52 |
| Walnut Shell | 741.9 | 1.41 | 1.03 | 1081 | 1.35 |
| Corn Cob | 1027 | 1.50 | 1.05 | 3135 | 1.05 |
| Aluminum Oxide | 182.4 | 1.49 | 1.11 | 82.8 | 3.77 |
| Quartz Sand 100-180 μm | 151.5 | 1.37 | 1.02 | 29.6 | 2.63 |
| Quartz Sand 180-200 μm | 190.3 | 1.06 | .99 | 50.9 | 2.63 |
| Quartz Sand 200-280 μm | 239.6 | 1.19 | .97 | 74.0 | 2.63 |
| Quartz Sand 280-300 μm | 289.8 | 1.04 | .99 | 108 | 2.63 |
| Quartz Sand 300-450 μm | 359.2 | 1.26 | 1.00 | 169 | 2.63 |
| Quartz Sand 450-500 μm | 473.1 | 1.05 | .99 | 315 | 2.63 |
| JSC-1 Lunar Simulant | 87.80 | 6.58 | 1.14 | 15.9 | 3.1 |

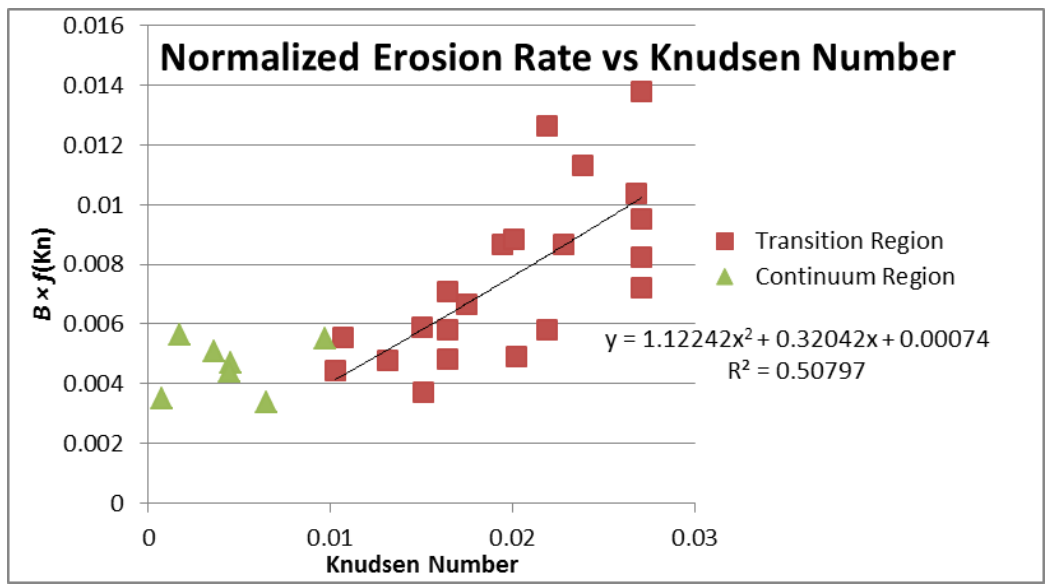

**Figure 3. Play Sand Test Cases.**

Ten cases with quartz of the various particle sizes were performed all at a chamber pressure of $P_0$ = 9.8 +/- 1 Torr (1293-1320 Pa) with two different velocities $v$=130 +/- 2 m/s and $v$=186 +/- 2 m/s representing two different jet dynamic pressures (nominally 275 Pa and 555 Pa). The particles' specific gravities are all the same so this isolates the effect of varying particle size (for the two different jet velocities). These results are shown in Figure 4.

---

[1] CU=Coefficient of Uniformity; CC=Coefficient of Curvature.

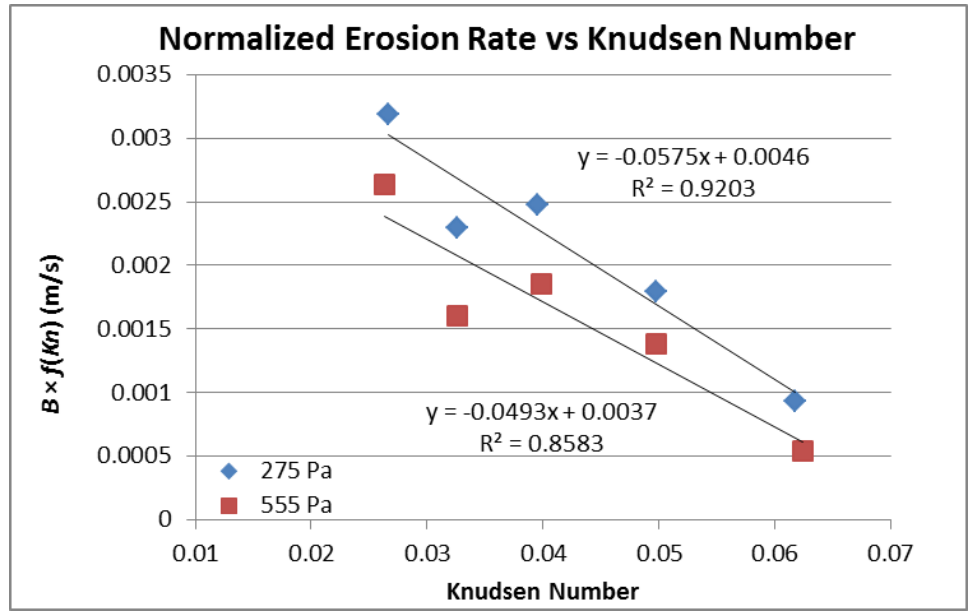

Figure 4. Quartz Sand Test Cases.

Two features are surprising. First, the data do not collapse to a single curve after normalizing by the continuum scaling, which shows the dependence on jet dynamic pressure is not the same in the transitional regime. Thus, the hypothesis stated at the beginning of the paper is disproven and $f$ must be more complicated than $f = f(Kn)$. Second, the erosion rate decreases with $Kn$ rather than increasing as in Figure 3. In that case $Kn$ was increasing because $P_0$ decreased while $D$ remained constant. Here, $Kn$ increased because $D$ decreased while $P_0$ remained constant. Since the erosion rate had been normalized by continuum scaling, the plot should not have shown a trend in $D$. This suggests that scaling with $D$ is different in the transitional regime than in the continuum regime.

Note also that the second data point from the top of Figure 4 is lower than the third data point (for both dynamic pressures) and thus appears to have been swapped. It is possible the quartz sand in the two test boxes were accidentally reversed. It is impossible to tell now since the equipment was disassembled from the PAL vacuum chamber. The data have also been plotted with these data points swapped (not shown here), which produced a better fit but had little effect on the overall trend lines.

Some of the test cases with a variety of media (glass beads, plastic beads, ground walnut shell, ground corn cob, or aluminum oxide powder) are shown in Figure 5. These were taken all at one chamber pressure $P_0 = 9.8$ Torr (1306 Pa). Both $D$ and $\varrho_g$ are varied with these materials. It seems unlikely that the scaling with respect to $\varrho_g$ differs between the continuum and transitional regimes since the gas flow around the grains knows nothing of the density of the grains. In Figure 5, the values of $\dot{V}$ are normalized as follows. First, each one is multiplied by $D^\alpha$ using particle size $D$ for the corresponding particle type, where the exponent $\alpha=1$ is the erosion scaling for the continuum regime. Thus, all normalized values of $\dot{V}$ should be constant if the continuum scaling in $D$ is the same in the transitional region. Second, each value of $\dot{V}$ was divided by the case with the lowest $Kn$ so that all the values would be in the neighborhood of unity on this plot for convenience. The results are plotted two ways, first with $\alpha = 1$ and then with $\alpha = -1$, which appears to be the correct scaling in the transitional regime as attested by the fact that this value of $\alpha$ does make the values of $\dot{V}$ about the same across all $Kn$ on the plot. This implies that erosion rate is proportional to $D$ in the transitional regime (although more data are needed to be sure), whereas it is inversely proportional to $D$ in the continuum regime. As a check, Figure 4 was re-plotted with this new scaling in $D$ (not shown here). It changes the slopes of the two linear trend lines to more nearly flat in agreement with $\alpha = -1$ in Figure 5.

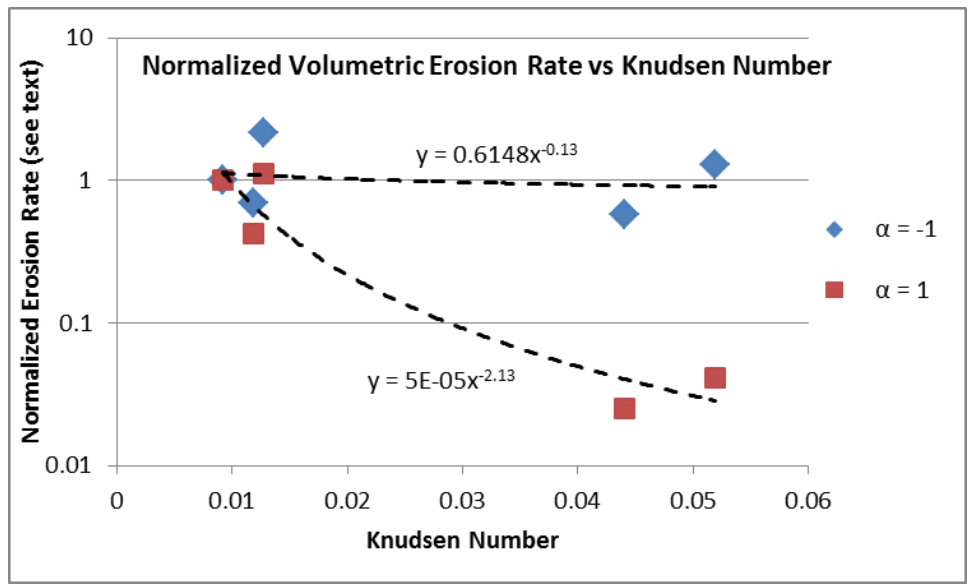

Figure 5. Mixed Media Test Cases

Partial results for the plastic beads are shown in Figure 6. This shows erosion rate (not normalized) versus gas velocity. All these cases were performed at nearly constant chamber pressure of 4.5 to 4.8 Torr (600 to 640 Pa) resulting in $Kn = 0.24$ to 0.26 in the transitional regime. A power law fit to these data show that erosion rate scales as velocity with an exponent of 2.4 in these cases. This differs from continuum scaling where the velocity exponent is 2.

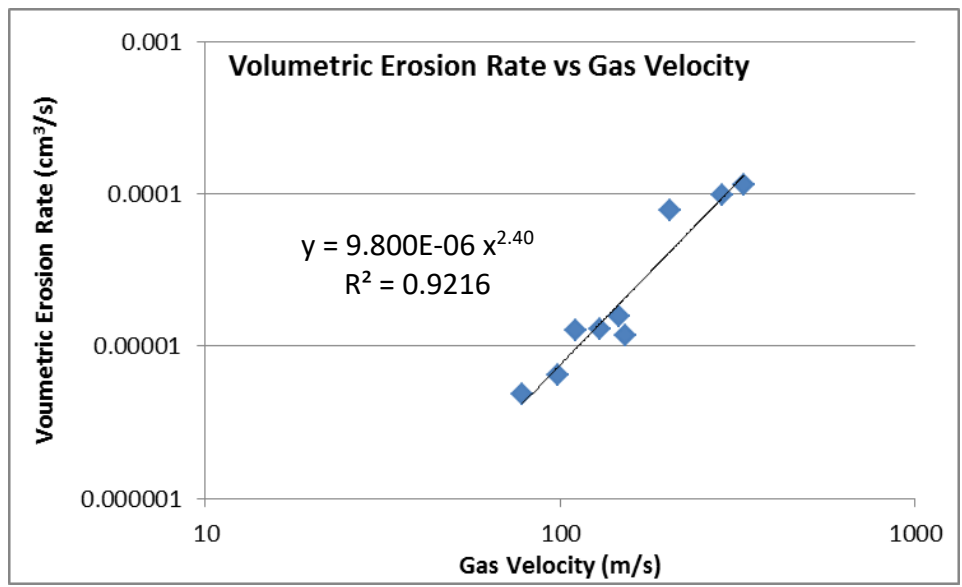

**Figure 6. Plastic Bead Test Cases**

Partial results for JSC-1A lunar soil simulant are shown in Figure 7. A power law fit to these data show that erosion rate scales as velocity with an exponent of 2.35 in these cases, in very good agreement with the exponent in Figure 6.

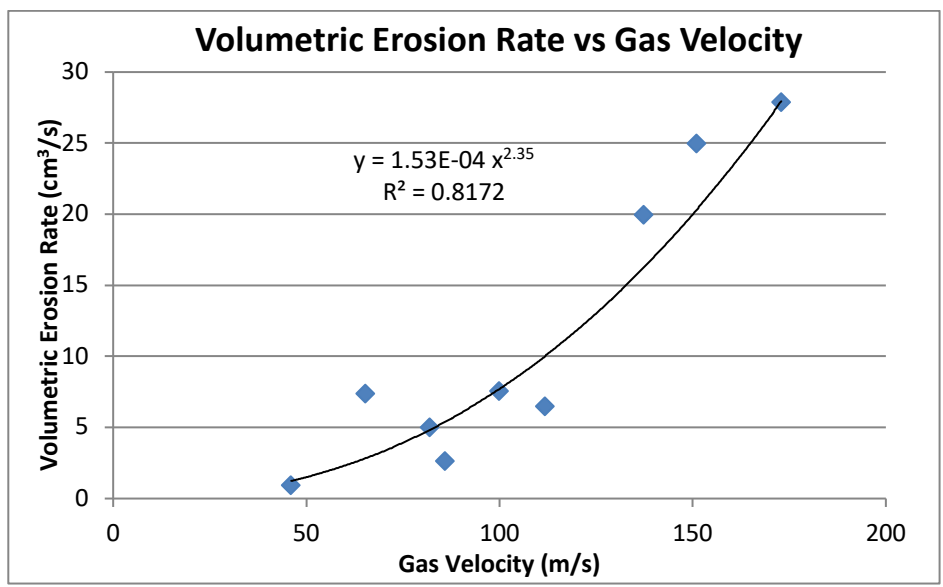

**Figure 7. JSC-1A Lunar Soil Simulant Test Cases**

The data set also includes more complicated combinations of parameter variations, but it is not clear yet how to reduce these into meaningful scaling relationships.

**DISCUSSION AND CONCLUSIONS**

Interaction of the gas and solid surface creates a boundary layer described by the roughness of the soil and the shear stress in the gas. Shear stress is established across the boundary layer by viscosity and by turbulent mixing. When gas is rarefied, the concept of viscosity breaks down because molecules travel long distances relative to the flow geometry without colliding with other gas molecules. Also, for rarefied conditions the spectrum of turbulence is truncated at larger diameter vortices, because gas molecules do not collide over short enough distances to close the loop on smaller vortices. Therefore, transmission of momentum across the boundary layer is modified. It seems likely that scaling of erosion rate in the transitional regime is more complicated than scaling in the continuum regime, which explains the difficulty in identifying simple scaling laws. It may be that there are multiple Knudsen numbers relevant to erosion physics. There is a Knudsen number relative to the diameter of a sand grain. There is another relative to the pipe exit diameter. Another is relative to the boundary layer thickness. It may be that scaling is a complex function of multiple Knudsen numbers. Also, preliminary results suggest the scaling with respect to gas density and velocity, if expressed as power laws, may have non-integer exponents. More progress analyzing this data set will be reported in future publications. So far it is apparent that scaling relationships in the transitional regime are not the same as in the continuum regime. More experimental work will be needed to really solve this problem.

**ACKNOWLEDGEMENT**

The author acknowledges and expresses gratitude for funding through NASA's Lunar Advanced Science and Exploration Research (LASER) program, award number NNX09AD07A, "Erosion and Transport of Lunar Regolith by the Exhaust Plumes of Landing Spacecraft."